\documentstyle[12pt,a4]{article}
\begin{document}
\title{New cosmological solutions and stability analysis in full extended
thermodynamics}
\author{Luis P. Chimento,  Alejandro S. Jakubi      \\
{\small Departamento de F\'{\i}sica, Facultad de Ciencias Exactas y Naturales
, }\\
{\small Universidad de Buenos Aires }\\
{\small  Ciudad  Universitaria,  Pabell\'{o}n  I,  1428 Buenos Aires, Argentina
.}\\
\\
Vicen\c c M\'endez \\
{\small Departament de F\'{\i}sica. Facultat de Ci\`encies, edifici Cc}\\
{\small  Universitat Aut\`onoma de Barcelona}\\
{\small E-08193 Bellaterra (Barcelona) Spain}}

\maketitle

\begin{abstract}
The Einstein's field equations of FRW universes filled with a
dissipative fluid described by full theory of causal transport equations are
analyzed. New exact solutions are found using a non-local transformations on
the nonlinear differential equation for the Hubble factor. The stability of
the de Sitter and asymptotically friedmannian solutions are analyzed using
Lyapunov function method.
\end{abstract}

\newpage

\section{Introduction}

Quantum effects played a important role in the early Universe. For instance,
vacuum polarisation terms and particle production arise from a quantum
description of matter.  Some other processes capable of producing important
dissipative stresses include interactions between matter and radiation
\cite{Wei71} and different components of dark matter \cite{Pav93}.
Phenomenologically, these processes may be modeled in terms of a classical
bulk viscosity \cite{Hu82}.

A relativistic second-order theory of non-equilibrium thermodynamics was
developed in  \cite{Isr76}. This formulation generalizes the expression of the
entropy flux by the inclusion of quadratic terms in dissipative
non-equilibrium magnitudes. Another formulation of this theory, called
Extended Irreversible Thermodynamics has been made in \cite{Pav82}
\cite{Jou93}. There, non-equilibrium magnitudes are introduced  and a
generalized Gibbs equation including the dissipative magnitudes is employed.

Recently, qualitative analysis of homogeneous cosmological models
using the full causal transport theory derived from the extended
irreversible thermodynamic \cite{cm}, \cite{vm} has been made to
investigate the stability of the particular solutions of the Einstein's
field equations. Several authors studied the stability of non flat
causal cosmological models using non usual equations of state
$p/\theta ^2 = p_0 x^l$, $\zeta /\theta = \zeta _0 x^m$ \cite{cm}.
They are able to get an autonomous system of differential equations
in a phase space $(x,y)$ where $x= 3\rho/ \theta ^2$ and $y=9\Pi
/\theta ^2$. Other authors \cite{vm} have studied flat models with
the widely used equations of state $p=(\gamma -1)\rho$ and $\zeta =
\alpha \rho ^m$. They have got an autonomous system in the phase space
$(H,\dot{H})$. In both studies, always the linearization technique
near stationary solutions in the phase space have been used.
Some other papers where the full version of the transport
equation has been used are \cite{His91} \cite{Zak93b} \cite{Rom}
\cite{Maa95}.

In this work, together with another one \cite{Chi96d}, new exact cosmological
solutions of full causal models are found and the asymptotic stability of
solutions present in non-causal and truncated causal models is investigated
using the direct method of Lyapunov. Few authors have used this method to
analyze the stability of causal models though it provides useful information
of the dynamical behaviour of the system, not only near the stationary
solutions but also far away from them  \cite{Chi96d} \cite{Chi93} \cite{Kra}.
In the present paper we assume that the cosmic fluid is described locally by
the equation of state $p=nT$, while in the paper
\cite{Chi96d} a power-law relationship between the equilibrium temperature and
energy density was assumed.

This work summarizes as follows. In section II we state the basic equations of
the model, they are solved in section III for $m=1/2$, the two-parameter
families of solutions are classified and their asymptotic behaviors are
analyzed. Also, the thermodynamics of these solutions is considered.  In
section IV we study the stability conditions for de Sitter and asymptotically
Friedman solutions for $m\neq 1/2$. Finally in section V we summarize the main
conclusions of the paper.

\section{Hydrodynamic description for the cosmic fluid. Basic equations}

For the fluid approach to remain valid, the mean interaction
time $t_c$ of the fluid particles must be much less than the
expansion rate. This requires, $t_c\ll H^{-1}$, where $H$ is the
Hubble expansion rate. If the fluid is in or close to equilibrium,
then this restriction ensures that the fluid has a well-defined local
temperature. For a cosmic fluid without shear viscosity or heat flow,
the energy momentum tensor is

\begin{equation}
T_{ab}=(\rho+p+\Pi)u_{a}u_{b}+(p+\Pi)g_{ab},
\label{eq:Tab}
\end{equation}

\noindent
where $u_a$ is the four velocity, $\rho$ the energy density, $p$ the
equilibrium pressure and $\Pi$ the bulk viscous pressure. In an
expanding fluid, the dissipation due to $\Pi$ leads to a decrease in
kinetic energy and therefore in pressure, so $\Pi \leq 0$. The
particle flow vector $N^a$ is given by $N^a=nu^a$ where $n$ is the
particle number density. In absence of vector and tensor dissipation,
the entropy four-flux will be of the form $S^a=S_{eff}n^a$ where
$S_{eff}$ is the effective non--equilibrium specific entropy which
must be positive. 

In the Israel/Stewart theory

\begin{equation}
S_{eff}=S-\frac{\tau}{2nT\zeta}\Pi ^2
\label{ise}\end{equation}
where $\tau$ is the relaxation time, $T$ is the local--equilibrium
temperature and $\zeta$ is the coefficient of bulk viscosity. The
conservation laws $N^a_{\;;a}=0$ and $T^{ab}_{\;;b}=0$ imply

\begin{equation}
\dot{n}+3Hn=0\;\;\mbox{and}\;\;\dot{\rho}+3H(\rho+p+\Pi)=0
\label{ceqs}\end{equation}
respectively, where $3H=u^{a}_{\;;a}$ is the fluid expansion and
$\dot{n} = n_{,a}u^a$. Combining equations (\ref{ceqs}) with the
Gibbs equation for the local--equilibrium variables $S$ and $T$
\begin{equation}
TdS = d\left(\frac{\rho}{n}\right) + p d\left(\frac{1}{n} \right)
\label{gie}\end{equation}
we get
\begin{equation}
\dot{S}= -\frac{3H\Pi}{nT}.
\label{dots}\end{equation}

From (\ref{ise}) and (\ref{ceqs}) we find
\begin{equation}
S^a_{\;;a}=-\frac{\Pi}{T}\left[3H+\frac{\tau}{\zeta}\dot{\Pi}+\frac{1}{2}\Pi
T\left(\frac{\tau u^a}{\zeta T} \right)_{;a} \right]
\label{enp}\end{equation} 
for the entropy production rate. The simplest way to guarantee the
second law of thermodynamics $S^a_{;a}\ge 0$ is establishing the
transport equation
\begin{eqnarray}
\Pi + \tau \dot{\Pi} = -3\zeta H - \frac{1}{2}\tau \Pi \left(
3H + \frac{\dot{\tau}}{\tau} -\frac{\dot{T}}{T} - \frac{\dot{\zeta}}{\zeta}
\right).
\label{eq:mc}
\end{eqnarray}
for $\Pi$, leading to $S^a_{;a}=\Pi ^2/\zeta T$.

We need to specify equations of state  $\rho = \rho (T,n)$ and $p = p(T,n)$
for the cosmological fluid. In order to have a simple model, we assume that
the cosmic viscous fluid may be locally described by an equilibrium  equation
of state that has the approximated form of an ideal gas and the $\gamma$-law:

\begin{eqnarray}\label{ideal}
p&=& nT\cr
\rho&=& \frac{1}{\gamma -1}nT . 
\end{eqnarray}
Note that both $T$ and $n$ are independent variables and only in the
equilibrium limit the particle number density depends exclusively on
$T$. 

From (\ref{ceqs}) and (\ref{ideal}) the expression for the
temperature is
\begin{equation}
T = \frac{\gamma -1}{n_0}R^3\rho .
\label{te1}\end{equation}  
We specialise to a flat FRW universe, where the field equations are

\begin{eqnarray}
H^2& =& \frac{1}{3}\rho ,\cr
3(\dot{H} + H^2)&=& -\frac{1}{2}(\rho +3p+3\Pi ),
\label{eq:ee}
\end{eqnarray}
where $H\equiv\dot{R}/R$ is the Hubble factor, $R(t)$ the cosmic scale factor
of the Robertson-Walker metric,  $8\pi G=c=1$ and
an overdot denotes differentiation respect to time $t$.
The relaxation time $\tau$ is related to the bulk viscosity by \cite{hanno}
\begin{equation}
\tau = \frac{\zeta}{v^2\gamma\rho}
\label{tau1}\end{equation}
where $v$ is the dissipative contribution to the speed of sound. By
causality $v^2\leq 2-\gamma$. The bulk viscosity coefficient is often
taken to be $\zeta = \alpha \rho ^m$ where $\alpha$ is a positive constant.

An other interesting restriction is that, since a dissipative
expansion is non--thermalising, the rate for some
interaction $(\sim \tau ^{-1})$, which is crucial to maintaining
equilibrium, must remain lower that the expansion rate $H$, so that
the fluid cannot adjust sufficiently rapidly to the changes in
temperature induced by expansion \cite{Maa95}.

Finally, using (\ref{eq:mc}), 
(\ref{te1}), (\ref{eq:ee}) and (\ref{tau1})
we get, for the evolution of $H$, the equation
\begin{equation}
H\ddot{H} + \frac{v^2\gamma}{\delta}\dot{H}H^{3-2m}-2\dot{H}
^2+\frac{3v^2\gamma H^4}{2}\left(\frac{\gamma}{\delta}H^{1-2m}-3 \right)=0
\label{eq:h}
\end{equation}
where $\delta =\alpha 3^{m-1}$, and we are assuming that the Universe
is expanding, that is $H>0$.
Using (\ref{dots}), (\ref{ideal}) and (\ref{eq:ee}) we get, after integrating
\begin{equation}
S(t)=S_0 + \frac{1}{\gamma -1}\ln (H^2R^{3\gamma}).
\label{dots1}\end{equation}

\section{Exact solution for $m=1/2$}

The Einstein's field equations of FRW universes filled with a dissipative
fluid described by both the truncated and non-truncated causal transport
equations can be linearized and solved for a power-law fluid when $m=1/2$
\cite {Chi96d} \cite{Chi93}. In what follows, we shall find the general
solution for a fluid that obeys (\ref{ideal}). In
this case setting $m=1/2$ in (\ref{eq:h}) we find

\begin{equation} \label{300}
H\ddot{H} + \frac{v^2\gamma}{\delta}\dot{H}H^{2}-2\dot{H}^2+
\frac{3aH^4}{2}\left(\frac{\gamma}{\delta}-3 \right)=0. 
\end{equation}

\noindent With the change of variables $H=1/s$, $t= \delta\sigma/v^2\gamma  $,
equation (\ref{300}) becomes into

\begin{equation}
\frac{d^2s}{d\sigma ^2}+\frac{1}{s} \frac{ds}{d\sigma}+ \frac{\mu}{s}=0.
\label{eq:s}
\end{equation}
where $\mu =(3\delta/2v^2\gamma)(3\delta-\gamma)$.

Equation (\ref{eq:s}) is a particular case of the most general equation
\begin{equation}
\ddot{s}+ \alpha_1 f(s)\dot{s}+\alpha_2 f(s)\int f(s) ds + \alpha_3
f(s) =0 .
\label{eq:f}
\end{equation}
which linearizes through the nonlocal transformation \cite{Chi93} \cite{Chi96}
\cite{Chi96b} \cite{Chi96c}
\begin{equation}
z= \int f(s) ds  \qquad \eta =\int f(s) dt
\label{eq:trans}
\end{equation}
becoming
\begin{equation}
z''+ \alpha_1 z'+\alpha_2 z + \alpha_3=0
\label{eq:z}
\end{equation}

\noindent where the $'$ indicates derivative with respect to the new variable
$\eta$. In our case $\alpha_1=1$, $\alpha_2=0$, $\alpha_3=\mu$, $f(s)=1/s$ and
the nonlocal transformation is then

\begin{equation}
z= \ln s  \qquad \eta = \frac{v^2\gamma}{\delta}\int\frac{dt}{s}
\label{eq:trans1}
\end{equation}

\noindent
Thus, integrating (\ref{eq:z})

\begin{equation}
z(\eta) = c_1 -\mu \eta - c_2 e^{-\eta}
\label{eq:z2}
\end{equation}

\noindent we obtain the general solution of (\ref{300}) in parametric form

\begin{equation}
H(\eta) = H_0\exp\left(\mu\eta + c_2 e^{-\eta}\right).
\label{eq:h1}
\end{equation}

\begin{equation}
\Delta t = \frac{\delta}{v^2\gamma H_0} \int d\eta
\exp\left({-\mu \eta - c_2 e^{-\eta}}\right).
\label{eq:sfo}
\end{equation}

\noindent where $\Delta t=t-t_0$, and  $H_0$, $c_2$, $t_0$ are
arbitrary integration constants. The scale factor can also be found in terms
of $\eta$ as

\begin{equation} R(\eta) = R_0 \exp\left({\frac{\delta}{v^2\gamma}\eta}\right)
\label{eq:r1}
\end{equation}
where $R_0$ is another integration constant. In this case $\zeta\sim
H$ and $\tau \sim H^{-1}$.

Below, we show explicitly exact solutions and the asymptotic
behaviour of the general solution.

\subsection{General solution for $\mu =1$}
In the particular case $\mu =1$, Equation (\ref{eq:f}) has an additional form
invariance symmetry. Then, introducing the transformation \cite{Chi96b}

\begin{equation} \label{301}
s^{-1} = \frac{\dot{R}}{R} = -\frac{\delta}{v^2\gamma}
\frac{w^{-1}}{\int w^{-1}dt}
\end{equation}

\noindent
into (\ref{eq:s}) we find $w''(\sigma) =0$ and $w(\sigma) =
c_1\sigma +c_2$. Finally, the two-parameter family of solutions is given by

\begin{equation} \label{302}
R(t) = R_0\left(c \pm \ln \mid \Delta t\mid \right) ^{-\delta /v^2\gamma} ,
\end{equation}

\noindent
where $\delta = \left(\gamma + \sqrt{\gamma ^2 +8v^2\gamma} \right)/6$, 
and $c$ is an
arbitrary integration constant.
There exist solutions that expand so quickly that the scalar curvature
diverges when the scale factor diverges and they become infinite in a
finite amount of proper time. Some of these singular solutions, those with
minus, have a finite time span and particle horizons. Also, we have the
time reversal of the above solutions. These with plus lead to a
shrinking universe. In both cases, the dynamical information is enough
in order to select the optimal physical solutions.  Non--thermalising condition
requires in this case $v<\frac{1}{\sqrt{3}}\sqrt{1+2/3\gamma}$.

\subsection{General solution for $\mu=0$}

We note first that in this case ($\gamma /3 = \delta $) 
equation (\ref{eq:h}) admits a de Sitter stable
solution $H=H_0$, where $H_0$ is arbitrary \cite{vm}. On the other hand, for
$H\neq H_0$, equation (\ref{eq:s}) can be integrated twice by quadratures

\begin{equation} \label{32}
\Delta t=-\frac{1}{3v^2}\int \frac{dH}{H^2}\left(\ln\frac{H}{H_0}\right)
^{-1}
\end{equation}

\noindent
and we find these two-parameter families of solutions in implicit form

\begin{equation} \label{33}
\Delta t=\frac{1}{3v^2H_0}\mbox{E}_1\left(\ln\frac{H}{H_0}\right)
\end{equation}

\begin{equation} \label{34}
\Delta t=\frac{1}{3v^2H_0}\mbox{E}_1\left[\left(\frac{R}{R_0}\right)^
{-3v^2}\right]
\end{equation}

\noindent
for $H/H_0>1$, and

\begin{equation} \label{35}
\Delta t=-\frac{1}{3v^2H_0}\mbox{Ei}\left(-\ln\frac{H}{H_0}\right)
\end{equation}

\begin{equation} \label{36}
\Delta t=-\frac{1}{3v^2H_0}\mbox{Ei}\left[\left(\frac{R}{R_0}\right)^
{-3v^2}\right]
\end{equation}

\noindent for $H/H_0<1$, and where $\mbox{E}_1$ and $\mbox{Ei}$ are 
exponential integrals.
Solutions of both families have an asymptotically de Sitter stage for large
times, while their respective initial behavior (singular or nonsingular)
follows those of the solutions with $\mu=1$. Non--thermalising
condition requires here $v<1/\sqrt{3}$ and the second law of
thermodynamics does not impose additional restrictions.

\subsection{Particular solution for $c_2=0$}

If $c_2=0$ in  (\ref{eq:z2}) then $H(\eta =0) = H_0$ and $R(\eta
=0)=R_0$ and we find that $H =2/\left(3(\gamma-3\delta) \Delta
t\right)$ and the scale factor is given by

\begin{equation} \label{30}
R(t) = R_0 \Delta t^{\beta}
\end{equation}

\noindent
where $\beta =2/3\left(\gamma-3\delta\right)$. \noindent This solution is also
found if we set $H=\beta / \Delta t$ in (\ref{300}). If $\beta >0$, that is,
for $\delta <\gamma /3$, the Universe starts from a Big-Bang singularity and
then expands, while for $\delta >\gamma /3$ the scale factor diverges in a
finite amount of proper time. For $\delta =\gamma /3$, we have $\mu =0$ and
$z=c_1$. In this case $R=R_0 e^{H_0 t}$ is the de Sitter solution.
Non--thermalising condition and $S_{eff}\geq 0$, at late times, require that
for $v<1/\sqrt{3}$, $\alpha$ must be in the range
$\sqrt{3}v^2\gamma<\alpha<\gamma/\sqrt{3}$.

For
$$v<\frac{1}{\sqrt{3}}\sqrt{1-\frac{2}{3\gamma}}\;\;\mbox{and}\;\;\frac{\gamma}{\sqrt{3}}
-\frac{2\sqrt{3}}{9}<\alpha<\frac{\gamma}{\sqrt{3}}$$

\noindent
there exist inflationary thermodynamically consistent power--law solutions
given by (\ref{30}).
For
$$\sqrt{3}v^2\gamma<\alpha<
\frac{\gamma}{\sqrt{3}}-\frac{2\sqrt{3}}{9}$$
non--inflationary friedmannian thermodynamically consistent 
solutions given by (\ref{30}) exist. We conclude that whenever

$$v<\frac{1}{\sqrt{3}}\sqrt{1-\frac{2}{3\gamma}}$$

\noindent
inflationary power--law and non--inflationary scenarios are
thermodynamically consistent.

For $\beta =1/2$, that is $\alpha = \sqrt{3}(3\gamma -4)/9$ for
$4/3<\gamma<2$, the universe behave as if were filled by a perfect
fluid and were radiation dominated. For $\beta =2/3$, that is,
$\alpha = \sqrt{3}(\gamma -1)/3$, the universe behave as if were filled 
by a perfect fluid and were matter dominated.

\subsection{Analysis of the solutions}

Here we analyze the behavior of the general solution (\ref{eq:h1})-
(\ref{eq:sfo}). We begin by studying its asymptotic behavior near the
singularities and for large times.

For $H\rightarrow \infty$, $s\rightarrow 0$ and $z\rightarrow-\infty$. This
occurs in two cases. If $\mu >0$, $z\simeq -\mu \eta$ for $\eta \rightarrow
+\infty$. Then, using (\ref{eq:trans1}), we find that $H\sim e^{\mu
\eta}=-1/(\mu \sigma)$. So, there is a two-parameter family of solutions with
an explosive singularity at a finite time whose leading behavior is given by
(\ref{30}). Also, $H$ diverges when $\eta\to -\infty$ and $c_2>0$. In this
case, $H\sim\exp\left(c_2e^{-\eta}\right)$ and there is a two-parameter family
of solutions representing a Universe beginning to from a Big-Bang singularity
as

\begin{equation} \label{350}
R(t) \sim \left|\ln \left|\Delta t\right| \right|^{-3v^2}
\;\; \mbox{for}\; \Delta t\to 0
\end{equation}

The 
opposite extreme is $H\rightarrow 0$. In this case $s\rightarrow \infty$ and
$z\rightarrow+\infty$. If $\mu<0$  we obtain $z\simeq \mid
\mu\mid\eta$ for $\eta \rightarrow +\infty$. Using the (\ref{eq:trans1})
again, we obtain $H= e^{-\mid \mu \mid \eta} = 1/(\mid \mu \mid \sigma )$. So
there exists a biparametric family of asymptotically Friedmann solutions with
leading behavior given by (\ref{30}). Furthermore, these solutions are stable,
and (\ref{30}) is an attractor.

\noindent
When $\eta \to -\infty$, two different behaviors occur depending on whether
$c_2< 0$ or $c_2=0$. In the first case, $H\sim\exp\left(c_2e^{-\eta}\right)$
and there is a two-parameter family of solutions representing a Universe
beginning to evolve as

\begin{equation} \label{35b}
R(t) \sim \left|\ln \left|\Delta t\right| \right|^{-3v^2}
\;\; \mbox{for}\; t\to -\infty
\end{equation}

\noindent
In the second case the solution is (\ref{30}).

The two-parameter families of solutions can be classified according to their
number of singularities.

There is a family of singularity-free solutions which begin at $t=-\infty$ and
their behavior in the far future is Friedmann or have a final de Sitter stage.
These solutions do no have particle horizons. 

There is a family of solutions that begin as in the previous case but end at
an explosive singularity. These solutions do not have particle horizons. Also
there is a family of solutions that begin at a Big-Bang singularity and have a
final Friedmann or de Sitter behavior.

Finally, there is a family of solutions with a finite time-span that begin at
a Big-Bang singularity and end at an explosive singularity.

\subsection{Thermal consequences}

Inserting the general solution (\ref{eq:h1}), (\ref{eq:sfo}) in
(\ref{te1}) 

\begin{equation} \label{t}
T = 3\frac{\gamma -1}{n_0}R_0^3H_0^2
\exp\left[\frac{3\delta}{v^2\gamma}\left(1+3\delta-\gamma\right)\eta+
2c_2e^{-\eta}\right]
\end{equation}

\noindent we can investigate the evolution of the temperature for the
different families of solutions studying the limits near the singularities and
for large times. For $t\to\infty$, the temperature vanishes provided that
$\alpha<\sqrt{3}(\gamma -1)/3$, so the scale factor grows asymptotically
slower than $t^{2/3}$. The above condition implies that
$\mu<-\left(\gamma-1\right)/\left(2\gamma v\right)<0$. This shows that,
although a power-law inflationary scenario is possible, the temperature grows
unboundly in the far future.  For evolutions with a singularity in the past,
we have a hot Big-Bang beginning of the Universe because the temperature
diverges at the singularity. On the other hand, for evolutions ending at an
explosive singularity, the temperature diverges as a negative power of the
proper time interval to the singularity. There are two two-parameter families
of solutions which have a cool beginning with $R=0$ in the far past. These
contain a set of solutions that have a cool ending provided
$\alpha<\sqrt{3}(\gamma -1)/3$, otherwise they have a hot final stage. With a
fine-tuning of the parameters which corresponds to solutions with the
asymptotical behavior $R\sim t^{2/3}$ when $t\to\infty$, $T\to 3R_0^3
H_0^2\left(\gamma -1\right)/ n_0 $. In this particular case, the solution
(\ref{30}) has a constant temperature.

\section{Stability analysis for $m\neq 1/2$}

We will make use of the method of the Lyapunov function \cite{Kra} to
investigate the asymptotic stability of the de Sitter and asymptotically
Friedmann solutions that occur in the noncausal and truncated causal models
\cite{Chi93}, as well as in the full causal model a with power-law equation of
state \cite{Chi96d}.

\subsection{ Stability of the de Sitter solution}

Equation (\ref{eq:h}) admits a de Sitter solution \cite{vm}

\begin{equation}
H_0 = \left(\frac{3\delta}{\gamma} \right)^{1/(1-2m)}\;\; (m\neq 1/2).
\label{eq:dss}
\end{equation}

\noindent To study its stability we make the change of
variable $H=s^{-1}$ in (\ref{eq:h}) which
becomes into

\begin{equation}
\ddot{s}+\frac{v^2\gamma}{\delta}s^{2m-2}\dot{s}-\frac{3v^2\gamma^2}
{2\delta}s^{2m-2}+\frac{9v^2\gamma}{2s}=0
\label{eq:sli}
\end{equation}
multiplying by $\dot{s}$ one may rewrite (\ref{eq:sli}) as
\begin{equation}
\frac{d}{dt}\left(\frac{1}{2}\dot{s} ^2+ V(s) \right) = D(s,\dot{s})
\label{eq:liape}
\end{equation}
where the term $\dot{s} ^2/2$ represents the kinetic energy term of
the system and $V(s)$ the potential energy term. The dissipative term
$D(s,\dot{s})$ is negative definite
for all $s$. One may calculate these terms and finds
\begin{eqnarray}
& &V(s) =
-\frac{3v^2\gamma^2}{2\delta}\frac{s^{2m-1}}{2m-1}+\frac{9v^2\gamma}{2}\ln s
\;,\;\;  \cr
& & D(s,\dot{s}) = -\frac{v^2\gamma}{\delta}\dot{s}^2s^{2m-2}. 
\label{eq:sist}
\end{eqnarray}

\noindent The potential $V(s)$ has a unique minimum for $s>0$ at
$s_0=H_0^{-1}$ for $m<1/2$, while for $m>1/2$ it becomes a maximum.
Thus we find that an exponential inflationary regime is asymptotically stable
for $t\to\infty$ for any initial condition $H>0$ provided that $m<1/2$ (node
or stable focus), but this regime becomes unstable for $m>1/2$
(saddle point). This result improves over previous studies based
in small perturbations about the solution (\ref{eq:dss}) \cite{vm}.

\subsection{Stability of the asymptotically Friedmann solution}

For $m>1/2$ it is easy to check that (\ref{eq:h}) admits a solution whose
leading term is $2/(3\gamma t)$ for $t\to\infty$. This suggests to
make  the change of variables

\begin{equation} \label{20}
H = \frac{1}{u(z)t},\qquad t^{2m-1} = z
\end{equation}

\noindent
in (\ref{eq:h}) which takes the form

\begin{equation}
\label{21}
{\frac{d}{dz}}\left[{\frac{1}{2}}{u^{\prime}}^{2}+U(u,z)\right]=D(u,u',z)
\end{equation}

\noindent
where $'\equiv d/dz$. We consider that $u$ lies in a neighbourhood of
$u_0\equiv 3\gamma/2$. Thus, when $z\to\infty$

\begin{equation} \label{22}
U(u,z)=\frac{v^2\gamma}{2\delta(2m-1)^2}\left(\frac{u^{2m}}{m} -
\frac{3\gamma u^{2m-1}}{2m-1}
\right)\frac{1}{z}+O\left(\frac{1}{z^2}\right)
\end{equation}

\noindent
for $m\neq 0$,

\begin{equation} \label{221}
U(u,z)=\frac{v^2\gamma}{\delta}\left(\ln\frac{u}{u_0}+\frac{3\gamma}{2u}\right)
\frac{1}{z}+O\left(\frac{1}{z^2}\right)
\end{equation}

\noindent
for $m=0$, and

\begin{equation} \label{23}
D\left(u,u',z\right)=-\frac{v^2\gamma}{\delta(2m-1)}u^{2m}u'^2+
O\left(\frac{1}{z}\right) .
\end{equation}

As $U(u,z)$ has a unique minimum at $u_0$ for any $m\neq 1/2$,
and $D(u,u',z)$ is negative definite for $m>1/2$, we find that solutions
with leading Friedmann behaviour $R\sim t^{2/(3\gamma)}$ when $t\to\infty$ are
asymptotically stable for $m>1/2$, but become unstable for $m<1/2$. This
behavior has already been found in the truncated theory and the power-law
model \cite{pbj}, \cite{Chi96d}.

\section{Conclusions}

A cosmological model for a causal viscous fluid with an equation of state like
that of an ideal gas is analyzed carefully. Our main result is that many
qualitative features of this model are similar to those of the full causal
model with a power-law equation of state for the temperature and the truncated
model.

Applying a non-local integral transformation to the nonlinear differential
equation for the expansion rate it linearizes, and we find the general
solution of this model for $m=1/2$. Due to viscous effects, these new exact
solutions contain a two-parameter family that avoid the singularity and do not
have particle horizons. This interesting evolution has already been found
in the full causal theory, with a power-law equation of
state for the equilibrium temperature \cite{Chi96d}.

Among the set of solutions for some values of the parameters we have found
explicit exact ones  that give the qualitatively behavior of parametric
solutions for $\mu>0$. Also, there are solutions that begin at a Big-Bang
singularity and may represent a power-law inflationary stage or a matter or
radiation dominated era.

For $m=1/2$, the splitting of the large time asymptotic behavior of solutions
in terms of sgn$(\gamma-3\delta)$ follows closely that of the power-law model
and is the same as in the truncated model, which in its turn resembles the
classification for the noncausal solutions. As in the power-law case, no
family of solutions with extrema is found, and this appears to be the main
consequence of the full theory. On the other hand, a logarithmic leading
behavior appears near the Big-Bang singularity and in the remote past. This
kind of evolution has not been observed previously for these models. We note
that the rate of expansion in this logarithmic Big-Bang is larger than in a
regular power-law one.

The asymptotic behavior of solutions for this model was analyzed using the
method of Lyapunov for $m\neq 1/2$. We have shown that friedmannian solutions
are an attractor set of the biparametric family of solutions for $m\ge1/2$,
while they are unstable for $m<1/2$. This information is not available via the
stability analysis by linearization \cite{vm}. So we have improved this
analysis, obtaining more information away from the stationary solutions. Also
we have found that an exponential inflationary stage is asymptotically stable
for $m\le 1/2$ while it is unstable for $m>1/2$. Here also, the same
structural behavior as in the power-law, truncated \cite{pbj} and non-causal
models is observed.

It has been conjectured that due to the use of causal truncated  or noncausal
theories, viscosity-driven exponential inflationary
solutions are spurious  \cite{His91}. There, a cosmological model containing a dissipative
Boltzmann gas is studied. We show  here that this conclusion does not
generalize to models  described by another equation of state.

\vskip 3cm
\noindent
{\bf Acknowledgments}

We thank to Diego Pav\'on for his interesting comments about the results of
this paper. One of us (LPCh) wants to acknowledge the hospitality of the
Departament de F\'{\i}sica, Facultat de Ci\`encies, of the
Universitat Aut\`onoma de Barcelona, where part of this work was done. This
was partially supported by project 3091/92 CONICET and by the
Direcci\'on General de Investigaci\'on Cient\'{\i}fica y T\'ecnica of
the Spanish Ministry of Education and Science under grant PB94-0718.

\end{document}